\pdfoutput=1
\RequirePackage{fixltx2e}
\documentclass[aps,reprint,floatfix]{revtex4-1}

\usepackage{amsmath,amsfonts,amssymb} %
\usepackage{mathtools} %
\usepackage{wasysym} %
\usepackage{bm} %
\usepackage[bbgreekl]{mathbbol} %
\usepackage{graphicx} %
\usepackage[dvipsnames,table]{xcolor} %
\usepackage{tabularx} %
\usepackage[acronym]{glossaries} %
\usepackage{braket} %
\usepackage{etoolbox}
\usepackage[binary-units=true,detect-weight=true,detect-inline-weight=math]{siunitx} %
\usepackage{fancyhdr} %
\usepackage{microtype} %
\usepackage{natmove}
\usepackage[byname]{smartref} %
\usepackage[breaklinks, %
colorlinks, linkcolor=Blue, citecolor=Blue, urlcolor=Blue]{hyperref} %
\usepackage[version=3]{mhchem}
\AtBeginDocument{\usepackage{booktabs}} 

\newcommand{\RR}{\mathbf{R}} %
\def\be{\begin{equation}} %
\def\ee{\end{equation}} %
\def\bea{\begin{eqnarray}} %
\def\eea{\end{eqnarray}} %
\DeclareMathOperator{\Var}{Var}

\def\MyTitle{Symmetry adaptation in quantum chemistry calculations on
  a quantum computer} %
\def\MyAuthora{Ilya G. Ryabinkin} %
\def\MyAuthorc{Scott N. Genin} %
\graphicspath{{pics/}}

\renewcommand{\arraystretch}{1.00}

\newcolumntype{Y}{>{\centering\arraybackslash}X}
\robustify\bfseries

\newacronym{QPE}{QPE}{quantum phase estimation} %
\newacronym{CVQE}{CVQE}{constrained variational quantum eigensolver} %
\newacronym{VQE}{VQE}{variational quantum eigensolver} %
\newacronym{UCC}{UCC}{unitary coupled-cluster} %
\newacronym{QCC}{QCC}{qubit coupled-cluster} %
\newacronym{FCI}{FCI}{full configurational interaction} %
\newacronym{CASCI}{CASCI}{complete active space configurational
  interaction} %
\newacronym{CAS}{CAS}{complete active space} %
\newacronym{JW}{JW}{Jordan--Wigner} %
\newacronym{BK}{BK}{Bravyi--Kitaev} %
\newacronym[longplural={degrees of freedom}, %
firstplural={degrees of freedom (DOF)}, plural={DOF}]{DOF}{DOF}{degree
  of freedom} %
\newacronym[longplural={equations of motion}, %
firstplural={equations of motion (EOM)}, %
plural={EOM}]{EOM}{EOM}{equation of motion} %
\newacronym{PES}{PES}{potential energy surface} %
\newacronym{CI}{CI}{configuration interaction} %
\newacronym{QMF}{QMF}{qubit mean-field} %
\newacronym{SQP}{SQP}{sequential quadratic programming} %
\newacronym{RHF}{RHF}{restricted Hartree--Fock}
\newacronym{MO}{MO}{molecular orbital}

\begin{document}

\title{\MyTitle}

\author{\MyAuthora{}} %

\author{\MyAuthorc{}} %
\affiliation{OTI Lumionics Inc., 100 College St. \#351, Toronto,
  Ontario\, M5G 1L5, Canada} %

\date{\today}

\begin{abstract}
  Quantum chemistry calculations on a quantum computer frequently
  suffer from symmetry breaking: the situation when a state of assumed
  spin and number of electrons is contaminated with contributions of
  undesired symmetry. The situation may even culminate in convergence
  to a state of completely unexpected symmetry, \emph{e.g.} that of
  for a neutral species while a cation was expected. Previously, the
  \gls{CVQE} approach was proposed to alleviate this problem
  [\href{http://dx.doi.org/10.1021/acs.jctc.8b00943}{Ryabinkin
    \emph{et al.} (2018), J. Chem. Theory Comput.
    DOI:10.1021/acs.jctc.8b00943}] here we analyze alternative, more
  robust solutions. In particular, we investigate how symmetry
  information can be incorporated directly into qubit Hamiltonians. We
  identify three essentially different techniques, the symmetry
  projection, spectral shift, and spectral reflection methods, which
  are all capable of solving the problem albeit at different
  computational cost, measured as the length of the resulting qubit
  operators. On the examples of \ce{LiH} and \ce{H_2O} molecules we
  show that the spectral shift method, which is equivalent to
  penalizing states of wrong symmetry, is the most efficient, followed
  by spectral reflection, and symmetry projection.
\end{abstract}

\glsresetall

\maketitle

\section{Introduction}

Recent attention to practical applications of quantum computers,
especially in quantum chemistry, revived interest in some fundamental
properties of electronic Hamiltonians. Each quantum chemical
calculation on a quantum computer begins with construction of the
second-quantized electronic Hamiltonian,
\begin{equation}
  \label{eq:qe_ham}
  \hat H = \sum_{pq} h_{pq}(\RR) {\hat a}^\dagger_p {\hat a}_q +
  \frac{1}{2}\sum_{pqrs} g_{pqrs}(\RR) {\hat a}^\dagger_p {\hat
    a}^\dagger_q {\hat a}_s {\hat a}_r,
\end{equation}
where ${\hat a}_p^\dagger$ (${\hat a}_p$) are fermionic creation
(annihilation) operators, $h_{pq}(\RR)$ and $g_{pqrs}(\RR)$ are one-
and two-electron integrals in a spin-orbital basis containing
$N_\text{so}$ functions, and $\mathbf{R} = (\mathbf{R}_1, \dots)$ are
positions of nuclei in a molecule treated as
parameters~\cite{Helgaker:2000}. $N_\text{so}$ is twice a number of
spatial orbitals that constitute the primary one-particle basis for
computations, and $N_\text{so} \ge n$, where $n$ is the number of
electrons in a physical system of interest, to satisfy the Pauli
exclusion principle.

One of main features of the fermionic Hamiltonian~\eqref{eq:qe_ham},
which is frequently overlooked, is that its eigenvectors are elements
of the \emph{Fock space}---a direct product of Hilbert spaces for 0,
1, up to $N_\text{so}$ fermions (electrons). In a non-relativistic
domain, however, $\hat H$ does \emph{not} couple states with different
number of electrons and commutes with the electron number operator
\begin{equation}
  \label{eq:Nop}
  \hat N = \sum_i \hat a_i^\dagger \hat a_i.
\end{equation}
One would expect, therefore, a block-diagonal structure of the
Hamiltonian matrix in the eigenbasis of $\hat N$, with individual
blocks that couple $n$-electron ($n = 0, 1, \dots, N_\text{so}$)
states only. It is quite surprising that, to the best of our
knowledge, the explicit form of those blocks has never been derived.
Even more, there is an old piece of wisdom stated
by~\citet{Kutzelnigg:1982/jcp/3081} that such operators are much more
complicated than $\hat H$ itself:
\begin{quotation}
  The Fock space Hamiltonian $H$ (without fixed particle number) has a
  much simpler structure (in terms of the basic one- and two-electron
  integrals) than its projection $H_n$ to an $n$-particle Hilbert
  space, and it is hence worthwhile to investigate the possibility of
  "diagonalizing" the Fock space Hamiltonian H directly, without -
  specifying the particle number $n$.
\end{quotation}
Traditional quantum chemistry works around this problem by employing a
many-electron basis set in the form of Slater determinants, which are
by construction the eigenfunctions of $\hat N$ corresponding to a
specific number of electrons.

The above problem might look like purely academic, but it becomes
practical in the realm of quantum computing. To embed an electronic
structure problem on a quantum computer, one must transform the
electronic Hamiltonian~\eqref{eq:qe_ham} to a qubit form:
\begin{equation}
  \label{eq:spin_ham}
  \hat H = \sum_I C_I(\RR)\,\hat P_I,
\end{equation}
where $C_I(\RR)$ are numerical coefficients derived from the one- and
two-electron integrals and $\hat P_I$ are ``Pauli words'',
\begin{equation}
  \label{eq:Pi}
  \hat P_I = \cdots \hat w_{1}^{(I)} \, \hat w_{0}^{(I)},
\end{equation}
products of Pauli operators acting on different qubits
$\hat w_i^{(I)}$. Each $\hat w_i^{(I)}$ is one of $\hat x$, $\hat y$,
or $\hat z$ Pauli operators for the $i^{\rm th}$ qubit; their
canonical images are the respective Pauli matrix $\sigma_x$,
$\sigma_y$, and $\sigma_z$. The required transformation may be chosen
from a list of conventional ones, such as
\gls{JW}~\cite{Jordan:1928/zphys/631, AspuruGuzik:2005/sci/1704} or
more recent \gls{BK}~\cite{Bravyi:2002/aph/210,
  Seeley:2012/jcp/224109, Tranter:2015/ijqc/1431,
  Setia:2017/ArXiv/1712.00446, Havlicek:2017/pra/032332}. The number
of qubits acted upon by the Hamiltonian~\eqref{eq:spin_ham} is equal
to the number of spin-orbitals $N_\text{so}$. 
Since the fermion-to-qubit transformations are isospectral, every
eigenstate of ferminonic $\hat H$ has a counterpart in the set of
eigenstates of qubit $\hat H$. However, there is no simple and
convenient basis in the multi-qubit space that is also the eigenbasis
of $\hat N$ operator. Thus, a general qubit transformation that may be
employed to find eigenstates of the qubit Hamiltonian, either single-
or multi-qubit, has a potential to mix states with different number of
electrons. As a result, if a user does not monitor the mean value of
$\hat N$, there is a risk to obtain a physically meaningless
result~\cite{Ryabinkin:2018/jctc/0-cnstr}. Even worse, since
eigenvectors of the qubit Hamiltonian live in the Hilbert space of
$N_\text{so}$ qubits, the \emph{na\"{\i}ve} application of the
variational method may lead to a state with the number of electron
other than desired or expected. For example, it is generally true that
the electronic energy of a cationic species \ce{M^{+}} is higher than
that of neutral \ce{M} and thus the ground electronic state of
\ce{M^{+}} is among the \emph{excited} states of the qubit
Hamiltonian~\eqref{eq:spin_ham}.

It is not that such a problem is entirely new for the quantum
chemistry community. There exists another, a closely related one---how
to construct the approximate solutions of the electronic structure
problem that are also eigenfunctions of the total spin-squared
operator $\hat S^2$. Since $\hat H$ commutes with $\hat S^2$ as well,
this problem is totally equivalent to the one above. The exact
eigenfunctions of $\hat S^2$ are much more difficult to
construct~\cite{Pauncz:1979} than Slater determinants, so that the
lack of a computationally convenient basis in both problems is a
common point.

In this paper we assess various methods for symmetry restoration in
quantum chemical calculations carried out on a quantum computer
(\emph{i.e.} in the qubit representation). We try to be
hardware-agnostic, aiming at applications on a generic universal
quantum computing device. At the same time, it is not possible to be
entirely method-independent. There are a couple of methods made for
quantum computers, and the one, which will of our primary concern, is
the \gls{VQE} method.

The most general description of the \gls{VQE} method is as follows.
First, a trial wave function is parametrized as
\begin{equation}
  \label{eq:VQE-wf-Ansatz}
  \ket{\Psi(\boldsymbol \kappa)} = \hat U(\boldsymbol \kappa) \ket{0},
\end{equation}
where $\ket{0}$ is the fixed initial state of a quantum register
(\emph{e.g.} all $N_\text{so}$ qubits in ``spin-up'' states) and
$\hat U(\boldsymbol \kappa)$ is a unitary transformation parametrized
by a set of externally controlled amplitudes
$\boldsymbol \kappa = \kappa_1, \dots$ implemented as a quantum
circuit. Running the quantum computer brings it to a state
$\ket{\Psi(\boldsymbol \kappa)}$, and the mean energy
\begin{align}
  \label{eq:E_est}
  E(\boldsymbol \kappa) & = \braket{\Psi(\boldsymbol \kappa) | \hat H
                          | \Psi(\boldsymbol \kappa)} \nonumber \\
                        & = \sum_I C_I \braket{\Psi(\boldsymbol
                          \kappa)| \hat P_I | \Psi(\boldsymbol \kappa)}
\end{align}
is calculated by weighting measurements of individual Pauli words with
the coefficients $C_I$ as in the second line of Eq.~\eqref{eq:E_est}.
Finally, a classical computer minimizes $E(\boldsymbol \kappa)$ with
respect to $\boldsymbol \kappa$ to produce a ground-state energy
estimate
\begin{equation}
  \label{eq:classical_min}
  E = \min_{\boldsymbol \kappa}  E(\boldsymbol \kappa) \ge E_0.
\end{equation}
As we argued above, the whole procedure might not necessarily respect
symmetry and lead to wrong values of $\braket{\hat N}$ and/or
$\braket{\hat S^2}$\cite{Ryabinkin:2018/jctc/0-cnstr}.

There are three places in the \gls{VQE} procedure where symmetry can
be enforced. First, the Hamiltonian itself can be transformed to
exclude or penalize states with undesired values of symmetry
operators. As long as states with the certain number of electrons are
concerned, this procedure is equivalent to the construction of
sub-blocks $\hat H_n$ of the original Fock space Hamiltonian, whose
existence have been alluded by Kutzelnigg. Different ways to
incorporate symmetry information into the Hamiltonian have direct
measure of efficacy: shorter expansions in Eq.~\eqref{eq:spin_ham}
have clear preference. Second, appropriate symmetry constrains might
be introduced to the quantum circuit by modifying or augmenting
$\hat U$; we do not discuss this perspective here leaving it for
further studies. Finally, the energy minimization in
Eq.~\eqref{eq:classical_min} can be supplemented with constraints for
the mean values of $\hat N$ or
$\hat S^2$~\cite{Ryabinkin:2018/jctc/0-cnstr}. Mean-value constraints
have an important advantage that they incur almost no \emph{quantum}
overhead; in other words, very few, if any, additional Pauli word
measurements apart from those that have already been done for energy
are needed, and both the Hamiltonian and the quantum circuit were left
untouched. On the other hand, constraining the mean value still allows
for non-zero \emph{fluctuations} around this mean, whose origins may
be explained as follows. Variational optimization of trial
states~\eqref{eq:VQE-wf-Ansatz}, generally, leads to a wave function
$\ket{\Psi(\boldsymbol \kappa_\text{opt})}$ that is an eigenfunction
of \emph{not} the original $\hat H$, but a different operator,
$\hat H'$. A good example of such an approximate Hamiltonian is the
zero-order (unperturbed) Hamiltonian of the M{\o}ller--Plesset
perturbation theory,
\begin{equation}
  \label{eq:H_0_fock}
  \hat H' \equiv \hat H^{(0)} = \sum_i \hat F(i),
\end{equation}
which is a sum of Fock operators for different electrons, see Sec.~10
of Ref.~\citenum{Helgaker:2000}. Approximate wave functions
$\ket{\Phi}$, which are Slater determinants, are the eigenfunctions of
$H^{(0)}$ but not $\hat H$; however they extremize the expectation
value of $\hat H$ in the space of determinantal wave functions,
\begin{equation}
  \label{eq:E_HF}
  E_\text{HF} = \min_{\Phi} \braket{\Phi|\hat H|\Phi} \ge E_0.
\end{equation}
In classical mechanics objects like $\hat H'$ are known as ``shadow
Hamiltonians''~\cite{Engle:2005/jcp/432} and play an important role in
construction of propagation algorithms (\emph{e.g.} the symplectic
ones) that conserve certain symmetries. In the context of this work it
is important that $\hat H'$ does \emph{not}, in generally, commute
with symmetry operators; hence, the common system of eigenfunction
does not exist, and the mean values of both energy and symmetry
operators may exhibit non-zero fluctuations. It might be enticing,
therefore, to constrain not only the mean-value but also the
\emph{variance} of symmetry operators at the minimization
stage~\eqref{eq:classical_min}, but as we show below, this is almost
equivalent to the direct modification of the electronic Hamiltonian
$\hat H$.

The rest of the paper is organized as follows. We compare various
methods of symmetry adaptation by the number of terms in the qubit
expansions [Eq.~\eqref{eq:spin_ham}] of the resulting qubit operators.
First of all, using a very direct but highly impractical algorithm, we
construct the qubit image of $\hat H_n$ for fixed $n$ corresponding to
a neutral molecule, whose existence was envisioned by Kutzelnigg. Thus
we can verify numerically his statement about the relative complexity
of this operator as compared to the (qubit image) of $\hat H$ itself.
$\hat H_n$ will be used as a reference point to compare complexity of
other operators. Then we consider more practical ways for building
symmetry-projected operators; the first will be the L\"owdin
projection technique~\cite{Lowdin:1964/rmp/966}. It turns out that the
L\"owdin projection technique can be interpreted as a special case of
spectral transformation. Having realized this, we discuss three
distinct ways of such transformation, namely: i) moving unwanted
states to 0, which is the transformation made by the L\"owdin
projection, ii) shifting them to high energy, and iii) reflecting them
through 0 to positive energies. We show that shifting is essentially
equivalent to penalizing the variance of the symmetry constraint,
while reflection has some connection to the Huzinaga's
equation~\cite{Huzinaga:1971/jcp/5543}. All the theoretical
considerations are supported by the numerical examples, in which all
introduced operators are explicitly constructed from fermionic
expressions for \ce{LiH} and \ce{H_2O} molecules.

\section{Theory}
\label{sec:theory}

\subsection{Symmetry projection by definition: A case of particle
  number-projected Hamiltonian}
\label{sec:part-numb-proj}

The electronic Fock-space Hamiltonian~\eqref{eq:qe_ham} as well as its
qubit image~\eqref{eq:spin_ham} are Hermitian operators that possess a
complete set of eigenfunctions. Thus, the following eigendecomposition
is valid:
\begin{align}
  \label{eq:H_eigendecomp}
  \hat H & = \sum_k E_k \ket{\Psi_k}\bra{\Psi_k} \nonumber \\
         & = \sum_k E_k(N_k, S_k) \ket{\Psi_k}\bra{\Psi_k}
\end{align}
where $E_k$ are the energy eigenvalues. The second line of
Eq.~\eqref{eq:H_eigendecomp} shows that symmetry labels $(N_k, S_k)$
can be attached to each energy level. $N_k$ and $S_k$ are the number
of electron and spin quantum numbers for the $k$-th level,
respectively. They are simultaneously measurable with the energy and
are related to the corresponding mean values as
$N_k = \braket{\Psi_k|\hat N|\Psi_k}$, $N_k = 0, 1 \ldots$ and
$S_k(S_k+1) = \braket{\Psi_k|\hat S^2|\Psi_k}$,
$S_k = 0, \frac{1}{2}, 1, \frac{3}{2} \dots$. An energy operator that
acts within, say, an $n$-electron subspace can be written as:
\begin{equation}
  \label{eq:H_n_def}
  H_n = \sum_{k | N_k = n} E_k(N_k, S_k) \ket{\Psi_k}\bra{\Psi_k}.
\end{equation}
Of course, Eq.~\eqref{eq:H_n_def} is a highly \emph{impractical} way
to construct $H_n$ since it requires full diagonalization of the
original $\hat H$---a task that is prohibitively difficult for any but
the smallest molecular systems. However, the direct comparison of the
number of terms in the qubit image of $\hat H_n$ to that for the
original $\hat H$ sets a reference for gauging the complexity of
various symmetry-projected Hamiltonians. 

\subsection{L\"owdin's symmetry projectors}
\label{sec:symmetry-projectors}

One of the oldest approaches to enforce proper symmetry in quantum
chemistry calculations, which is still under
development~\cite{Pons:2018/ijqc/0}, is the projection technique due
to~\citet{Lowdin:1964/rmp/966}. A projection operator to a subspace of
functions that correspond to an eigenvalue $a_i$ of a given operator
$\hat A$ is explicitly constructed as:
\begin{equation}
  \label{eq:Lowdin-proj}
  \hat P_{a_i} = \prod_{j \ne i } \frac{ \hat A - a_j}{a_i - a_j}.
\end{equation}
Here $a_j \ne a_i$ are other eigenvalues of $\hat A$ which all have to
be known in advance. For both $\hat N$ and $\hat S^2$ eigenvalues are
known, see Sec.~\ref{sec:part-numb-proj}. Once the projector is
constructed, it can be directly applied to basis functions or used to
``dress'' the original Hamiltonian
\begin{equation}
  \label{eq:projected_H}
  \hat H_{a_i} = \hat P_{a_i} \hat H \hat P_{a_i}
\end{equation}
to obtain a new operator that acts within the desired subspace of
eigenfunctions of $\hat A$. Eqs.~(\ref{eq:Lowdin-proj}) and
\eqref{eq:projected_H} are valid for any $\hat A$, not necessarily the
symmetry operators; however, if $\hat A$ is a symmetry operator for
$\hat H$, in other words, if $\hat A$ and $\hat H$ commute,
$[\hat A, \hat H] = 0$, then Eq.~\eqref{eq:projected_H} can be
simplified to
\begin{align}
  \label{eq:projected_H_simp}
  \hat H_{a_i} & =  \hat P_{a_i} \hat H \hat P_{a_i} =  \hat H \hat
                 P_{a_i}^2 = \hat H \hat P_{a_i},
\end{align}
as $\hat P_{a_i}^2 = \hat P_{a_i}$ (idempotency condition).

It was quickly realized that as written, the
projector~\eqref{eq:Lowdin-proj} is a complicated many-electron
operator, and its explicit form is intractable. In practical
applications, the expression~\eqref{eq:Lowdin-proj} is simplified so
that to retain a single factor containing a nearest-neighbor
eigenvalue $a_{i+1}$ only~\cite{Baker:1989/jcp/1789}. This approach is
not without issues though~\cite{Koga:1991/cpl/359}. $\hat P_{a_i}$ is
an idempotent operator only if \emph{all} eigenvalues of $\hat A$ are
included in the product; a truncated expression is no longer a
projector, and projected energy calculated with the
operator~\eqref{eq:projected_H_simp} can fall below the exact
value~\cite{Koga:1991/cpl/359}.

We note that if $\hat A = \hat N$, the corresponding $n$-electron
number-projected Hamiltonian $P_n H P_n$ might coincide with the
operator~\eqref{eq:H_n_def}. Indeed, those two operators have
identical spectra; however, arbitrary rotations are allowed in
degenerate subspaces, so that the number of terms might be different
as for equivalent operators written in different basises. We verify
this assertion numerically in Sec.~\ref{sec:numerical-examples}.

The L\"owdin projection with the $\hat N$ operator has been recently
used in attempts to reduce the qubit size of some small
Hamiltonians~\cite{Moll:2016/jpa/295301}. Unfortunately, the proposed
procedure involved direct inspection of the tensor form of
intermediate operators, which is apparently not a scalable approach. A
subsequent generalization using the Clifford group
theory~\cite{Bravyi:2017/ArXiv/1701.08213} was able to exploit the
number-of-particle symmetry in a systematic manner, albeit only
partially, by separating sub-spaces with even/odd particle numbers.
Therefore, a question whether the L\"owdin projection technique may
provide systematic qubit reduction beyond certain $Z_2$ symmetries,
like the even/odd number of electrons, remains open.

\subsection{Penalty method as a spectral transformation}
\label{sec:penalty-method-as}

Whenever the projected operator, $P_{a_i} \hat H P_{a_i}$, is applied
to a wave function, it \emph{shifts} all unwanted contributions to 0
eigenvalue, but not \emph{eliminates}
them. 
This consideration shows that other spectral transformations, which
also move away unwanted states but are simpler than symmetry
projectors, may exist. Indeed, such transformations have already been
suggested in literature in the context of quantum
computing~\cite{Mcclean:2016/njp/023023, mcclean:2017/pra/042308}. One
can add a penalty term to the Hamiltonian~\eqref{eq:spin_ham} to
introduce a Lagrangian
\begin{equation}
  \label{eq:symm_constr}
  \hat{\mathcal{L}}_{\mu,\,a_i} = \hat H + \frac{\mu}{2} (\hat A - a_i)^2.
\end{equation}
$\mu > 0$ is a penalty parameter, a large fixed number and $1/2$ is
introduced for convenience. Here we consider a case of only one
symmetry constraint, as the generalization to many of them is
straightforward. Let $\ket{\Psi_k}$ be simultaneous eigenstate of
$\hat H$ and $\hat A$ corresponding to the eigenvalue $a_k \ne a_i$.
Then,
\begin{align}
  \label{eq:L_action}
  \hat{\mathcal{L}}_{\mu,\,a_i} \ket{\Psi_k}
  & = \hat H \ket{\Psi_k} + \frac{\mu}{2} (\hat A - a_i)^2 \ket{\Psi_k} \nonumber \\
  & = \left[E_k + \frac{\mu}{2} (a_k - a_i)^2\right] \ket{\Psi_k}.
\end{align}
It is clear that the state $\ket{\Psi_k}$ is now shifted upward in
energy by $\mu(a_k - a_i)^2/2 >0$. By increasing $\mu$, \emph{all}
undesired states can be pushed above 0. We call this process a
``spectral shift'' as opposed to a ``move to 0'' that is done by the
symmetry projector.

It must also be noted that the spectral shifting is ``almost''
equivalent to penalizing the variance of a symmetry operator. Indeed,
if we write the action of $\hat{\mathcal{L}}_{\mu,\,a_i}$ onto a
general normalized wave function $\ket{\Psi}$, not necessarily an
eigenfunction, and project to $\bra{\Psi}$, we obtain:
\begin{align}
  \label{eq:barL_ai}
  \braket{\Psi |\hat{\mathcal{L}}_{\mu,\,a_i}| \Psi}
  & = \braket{\Psi | \hat H | \Psi} +  \frac{\mu}{2} \braket{\Psi |
    (\hat A - a_i)^2 |\Psi}  \nonumber \\ 
  & = \braket{\hat H} + \frac{\mu}{2}\braket{\Psi | \hat A^2 -
    \braket{\hat A}^2 | \Psi} + \frac{\mu}{2}(\braket{\hat A} - a_i)^2
    \nonumber \\ 
  & = \braket{\hat H} + \frac{\mu}{2}\Var{\hat A} +
    \frac{\mu}{2}(\braket{\hat A} - a_i)^2,
\end{align}
where $\braket{\hat A}$ is a shortcut for
$\braket{\Psi | \hat A | \Psi}$. If a problem is feasible in a sense
that $\braket{\hat A} = a_i$, then penalizing the variance
$\Var{\hat A} = \braket{\Psi | \hat A^2 - \braket{\hat A}^2 | \Psi}$
by $\mu$ is entirely equivalent to work with the spectrally shifted
operator~\eqref{eq:symm_constr} in unconstrained
minimization~\eqref{eq:classical_min}.


\subsection{Huzinaga-style transformation}
\label{sec:huzin-style-transf}

The spectral shifting technique discused in
Sec.~\ref{sec:penalty-method-as} has a minor disadvantage of being
dependent on an arbitrarily chosen penalty parameter $\mu$. We show
below how to define a parameter-free spectral transformation that
leaves only desired states in the negative part of the energy
spectrum. Let us take a case of $\hat S^2$ as the first example. Since
$\hat H$ and $\hat S^2$ operators commute, they possess a common set
of eigenfunctions. Consider the action of $\hat S^2$ on the
eigendecomposition of $\hat H$, Eq.~\eqref{eq:H_eigendecomp}:
\begin{equation}
  \label{eq:S2_action}
  \hat S^2 \hat H = \sum_i S_i(S_i+1)E_i
  \ket{\Psi_i}\bra{\Psi_i},\quad S = 0, \frac{1}{2}, \dots
\end{equation}
Form an operator
\begin{equation}
  \label{eq:H_S}
  \hat{\mathcal{H}}_{S =0} = \hat H - \hat H {\hat S}^2 - {\hat S}^2\hat H,
\end{equation}
whose eigendecomposition reads:
\begin{align}
  \label{eq:H_S_eigendecomp}
  \hat{\mathcal{H}}_{S =0} & = \sum_{i,\, S = 0} E_i \ket{\Psi_i}\bra{\Psi_i}
                             \nonumber \\
                           & + \sum_{i,\, S\ne 0} -E_i\left[2S_i(S_i+1) -
                             1\right] \ket{\Psi_i}\bra{\Psi_i}. 
\end{align}
Note that $2S_i(S_i+1) - 1 > 0$ for $S \ge 1/2$, so that the
eigenstates of the Hamiltonian that are not singlets are shifted to
\emph{positive} energies since $-E_j >0$ for the ground- and low-lying
excited states. Thus, undesired states are moved above zero in the
spectrum. Note that the operator ~\eqref{eq:H_S} is linear in
$\hat S^2$ regardless of the size of a system, which is the main
advantage over the expression~\eqref{eq:projected_H} with a system
size-dependent projector~\eqref{eq:Lowdin-proj}.

Eq.~\eqref{eq:H_S} works because $2S_i(S_i+1) - 1 > 0$ for non-singlet
states; it will fail, for example, for triplet states as the singlet
ones will be pushed \emph{down} in this case. However, it is always
possible to form at most \emph{quadratic} in $\hat A$ operator
($\hat A$ must be a symmetry operator though), namely,
\begin{equation}
  \label{eq:H_A_decomp}
  \hat{\mathcal{H}}_{a_i} = \hat H - \hat H(\hat A - a_i)^2 - (\hat A -
  a_i)^2\hat H,
\end{equation}
which shares the same egenvectors with $\hat H$ in the subspace
corresponding to the target eigenvalue $a_i$, while other states are
pushed to the positive part of a spectrum. We intentionally chose a
symmetrized form of $\hat{\mathcal{H}}_{a_i}$ in
Eq.~\eqref{eq:H_A_decomp} to emphasize the similarity of our approach
with that of \citet{Huzinaga:1971/jcp/5543}. In their method a
similarly looking expression is used to find out an energy operator
that commutes with the projector on a set of given functions. If those
functions are the eigenfunctions of the original Hamiltonian, than
their method is equivalent to ours with the important distinction that
all undesired states are \emph{reflected} ($E_j \to -E_j$) with
respect to zero of energy.

\begin{table*}
  \centering
  \caption{Construction and properties of operators used in this work}
  \begin{tabularx}{1.0\textwidth}{>{}Xcc}
    \toprule
    Property                 & \multicolumn{2}{c}{Molecule} \\ \cmidrule{2-3}
                             & \ce{LiH}                               & \ce{H_2O} \\ \midrule
    \multicolumn{3}{X}{\bfseries Molecular parameters/Electronic structure calculations details} \\
    Molecular configuration  & $R(\ce{Li-H}) = \SI{3.20}{\angstrom}$  & $ R(\ce{O-H}) = \SI{2.05}{\angstrom}$  \\
                             &                                        & $\angle \ce{HOH} = \SI{107.6}{\degree}$ \\
    Nuclear-nuclear repulsion energy $V_{nn}$, \si{\hartree} & \num{0.496104}  & \num{4.290107} \\
    Atomic basis set\footnotemark[1]         & STO-3G                                 & 6-31G \\
    Total number of orbitals & 6 & 13 \\
    Molecular orbital set    &                       \multicolumn{2}{c}{Hartree--Fock (canonical)} \\
    Active space to generate fermionic $\hat H$ ($C_{2v}$ labelling)  & $2a_1$, $3a_1$, $4a_1$ & $1b_1$, $3a_1$, $4a_1$, $2b_1$ \\
    Number of electrons
    in active space, $n$          & 2                                     & 4 \\
    Number of spin-orbitals
    in active space, $N_\text{so}$ & 6                                & 8 \\
    Fock space dimension, $2^{N_\text{so}}$ & 64                        & 256 \\
    Sizes of $\hat N$ subspaces for $N = 0, 1, \dots,
    N_\text{so}$ &  (1, 6, 15, 20, 15, 6, 1)  & (1, 8, 28, 56, 70, 56, 28, 8, 1) \\
    Sizes of $\hat S^2$ subspaces for $S = 0, \frac{1}{2}, 1, \frac{3}{2}, \dots,
    \frac{N_\text{so}}{4}$ &  (14, 28, 18, 4)  & (42, 96, 81, 32, 5) \\[1ex]
    \multicolumn{3}{X}{\bfseries Qubit-space quantites} \\
    Fermion-to-qubit mapping & parity\footnotemark[2] & \acrfull{BK}\\
    Number of qubits & \multicolumn{2}{c}{$ = N_\text{so}$} \\
    Number of Pauli terms in qubit $\hat H$   & 118 & 185 \\
    Number of Pauli terms in qubit $\hat N$   & 7 &  9  \\
    Number of Pauli terms in qubit $\hat S^2$ & 40 & 77 \\
    \bottomrule
  \end{tabularx}
  \footnotetext[1]{From the Basis Set Exchange library~\cite{emsl-2}.}
  \footnotetext[2]{Described in
    Ref.~\citenum{Nielsen:2005/scholar_text}.}
  \label{tab:op_prop}
\end{table*}

There is one possible issue with this method. It is known that
negative-energy eigenstates of first-quantized atomic and molecular
Hamiltonians do not constitute a complete set. Already for the
simplest quantum system, the hydrogen atom, one should also include
positive-energy (scattering) eigensolutions~\cite{Landay_III}. Upon
spectral reflection those states may be brought into the
negative-energy manifold. We do not expect, however, it is a serious
problem for the following reason. The \emph{second-quantized}
Hamiltonian contains only a few, if any, positive-energy solutions
because in the most cases the set of molecular spin-orbitals that
determine the form~\eqref{eq:qe_ham} is chosen to describe low-energy
bound states, and bound and scattering states are very different in
their asymptotic behaviour. If, nonetheless, a molecular basis set
contains highly localized (tight) basis functions, they may have
substantial overlap with scattering states. In this case we recommend
to drop the highest-energy molecular spin-orbitals in the construction
of the Hamiltonian~\eqref{eq:qe_ham}. Also we recommend to choose the
zero of electronic energy for the Hamiltonian~\eqref{eq:qe_ham}
without accounting for the nuclear-nuclear repulsion energy---in that
case even highly-compressed molecular configurations will not have
electronic states with formally positive energies.

\section{Numerical examples}
\label{sec:numerical-examples}

We illustrate our developments on fermionic Hamiltonians and their
qubit images for \ce{LiH} and \ce{H_2O} molecules. The detailed
procedure for their construction is described in
Ref.~\citenum{Ryabinkin:2018/jctc/0}, however, essential information
about preparatory calculations is collected in
Table~\ref{tab:op_prop}. In contrast to
Ref.~\citenum{Ryabinkin:2018/jctc/0} we do not not exploit any qubit
symmetries, and our fermionic and qubit Hamiltonians are truly
isospectral. Thus, any qualitative statements made for the qubit
images remain valid for their fermionic originals.

\subsection{Number-projected Hamiltoninans}
\label{sec:numb-proj-hamilt}

First, we reconstructed the qubit images of the number-projected
operators from the sum-over-states expression, Eq.~\eqref{eq:H_n_def},
choosing $n = 2$ and 4 (neutral species) for \ce{LiH} and \ce{H2O},
respectively. The resulting operators contained \num{400} and
\num{1504} Pauli terms, which should be contrasted to \num{118} and
\num{185} terms in the qubit images of the fermionic Hamiltonians (see
Tables~\ref{tab:op_prop} and \ref{tab:op_size}). Thus, the statement
made by \citet{Kutzelnigg:1982/jcp/3081} is correct: the
number-projected operators are much lengthier than their Fock-space
originals. Second, we confirmed the numerical identity of the
number-projected operators computed by the
definition~\eqref{eq:H_n_def} with those computed by
Eq.~\eqref{eq:projected_H} using the L\"owdin
projector~\eqref{eq:Lowdin-proj} where $\hat A = \hat N$ and $a_i = n$
were chosen.

\begingroup %
\setlength{\tabcolsep}{0pt} \aboverulesep=0.1ex \belowrulesep=0.1ex
\begin{table*}
  \centering
  \caption{Number of terms in various symmetry-adapted operators.}
  \begin{tabularx}{1.0\textwidth}{>{}lYYY|YYY}
    \toprule
    Symmetry  & \multicolumn{6}{c}{Term count} \\
              & \multicolumn{3}{c}{\ce{LiH}}
              & \multicolumn{3}{c}{\ce{H_2O}} \\ \cmidrule{2-4} \cmidrule{5-7}
              & $\hat P \hat H \hat P$, Eq.~\eqref{eq:projected_H}
              & $\hat{\mathcal{L}}$, Eq.~\eqref{eq:symm_constr}
              & $\hat{\mathcal{H}}$, Eq.\,~\eqref{eq:H_A_decomp}
              & $\hat P \hat H \hat P$, Eq.~\eqref{eq:projected_H}
              & $\hat{\mathcal{L}}$, Eq.~\eqref{eq:symm_constr}
              & $\hat{\mathcal{H}}$, Eq.\,~\eqref{eq:H_A_decomp} \\  \midrule
    \multicolumn{1}{l}{\bfseries Electron number} \\
    Neutral & \cellcolor{Red!25}    400 & \cellcolor{Green!25} 118 & \cellcolor{Yellow!25} 381 & \cellcolor{Red!25} 1504 & \cellcolor{Green!25} 185 & \cellcolor{Yellow!25} 1143 \\
    Cation  & \cellcolor{Yellow!25} 248 & \cellcolor{Green!25} 118 & \cellcolor{Red!25}    381 & \cellcolor{Red!25} 1672 & \cellcolor{Green!25} 185 & \cellcolor{Yellow!25} 1511 \\
    Anion   & \cellcolor{Red!25}    320 & \cellcolor{Green!25} 118 & \cellcolor{Yellow!25} 273 & \cellcolor{Red!25} 1672 & \cellcolor{Green!25} 185 & \cellcolor{Yellow!25} 1511 \\[0.5ex]
    \multicolumn{1}{l}{\bfseries Spin} \\
    Singlet & \cellcolor{Red!25} 544 & \cellcolor{Green!25} 169 & \cellcolor{Yellow!25} 525\footnotemark[1] & \cellcolor{Red!25} 3216 & \cellcolor{Green!25} 695 & \cellcolor{Yellow!25} 2387\footnotemark[1]\\
    Triplet & \cellcolor{Red!25} 544 & \cellcolor{Green!25} 169 & \cellcolor{Red!25}    544 & \cellcolor{Red!25} 3216 & \cellcolor{Green!25} 695 & \cellcolor{Yellow!25} 3199 \\
    \bottomrule
  \end{tabularx}
  \footnotemark[1]{Singlet-state adaptation by a simplified formula,
    Eq.~\eqref{eq:H_S}.}
  \label{tab:op_size}
\end{table*}
\endgroup %

The number operator is also used to illustrate the differences in
spectra of the original $\hat H$ and its number operator-transformed
counterparts, Eqs.~\eqref{eq:projected_H}, \eqref{eq:symm_constr}, and
\eqref{eq:H_A_decomp}. Table~\ref{tab:spectra_comp} shows the energy
ordering and symmetry labels of the original levels and their matches
in the transformed spectra. As expected, all target levels are moved
by all three transformations to the bottom of the spectra, but the
energies and ordering of undesired states are different. The projected
Hamiltonian~\eqref{eq:projected_H} lumps all unwanted states at 0, the
shifted operator~\eqref{eq:symm_constr} moves them to the top, and
finally, the reflection (Huzinaga-style)
operator~\eqref{eq:H_A_decomp} makes all of them strongly positive in
energy. We note that appropriate choice of the penalty parameter $\mu$
can make the spectra of shifted and reflected operators similar.

\subsection{Complexity of various symmtetry-transformed operators}
\label{sec:compl-vari-symmt}

The number of terms in operators' expressions are collected in
Table~\ref{tab:op_size}. The trend is quite persistent with minor
deviations: The shortest expansion is given by the penalized
Hamiltonian, Eq.~\eqref{eq:symm_constr}, the longest -- by the
L\"owdin projection via Eq.~\eqref{eq:projected_H_simp}. In many
cases, especially when $\hat N$ adaptation is considered, the number
of terms in Eq.~\eqref{eq:symm_constr} is equal to that of the
original Hamiltonian; for the spin adaptation the increase is the
smallest.

This is apparently related to the fermionic rank of involved
operators: $\hat N$ is a single-particle operator and squaring it
makes it no more complex than $\hat H$, which is itself a two-particle
operator. Even the special case of a singlet projection in the
Huzinaga-style transformation, Eq.~\eqref{eq:H_S}, is not competitive
with the penalty method despite the formal four-particle nature of
products $\hat H \hat S^2$ and $ \hat S^2 \hat H$ in
Eq.~\eqref{eq:H_S} and the penalty term $\mu [\hat S^2 - 0]^2/2$ in
Eq.~\eqref{eq:symm_constr}.

\section{Conclusions}
\label{sec:conclusions}

We have introduced and analyzed three distinct ways, the L\"owding
symmetry projection, Eq.~\eqref{eq:projected_H_simp}, the spectral
shift, Eq.~\eqref{eq:symm_constr}, and the spectral reflection
(Huzinaga-style transformation), Eq.~\eqref{eq:H_A_decomp}, of
incorporating information about exact symmetries, namely,
particle-number and spin, directly into qubit Hamiltonians. The main
motivation was to relieve the computational burden put on a quantum
computer when states of certain symmetry are queried---the
symmetry-adapted energy operators greatly diminish odds for symmetry
breaking. The main challenge behind this approach is how to minimize
the size of the resulting operators. Direct calculations have shown
that it was indeed the problem: the symmetry-adapted operators were
lengthier than the originals, see Table~\ref{tab:op_size}. The clear
winner is the spectral shifting method~\eqref{eq:symm_constr}, which
in all cases provided the shortest expansion. For the case of the
number projection it is especially convenient: the transformed energy
operators have the same length as the originals, which we attributed
to the one-particle nature of the corresponding symmetry, $\hat N$.
The spin projection is more complicated, and in general one should
expect considerable expansion of the corresponding qubit operators. To
our surprise, the Huzinaga-style transformation,
Eq.~\eqref{eq:H_A_decomp} is not competitive with the spectral shift
method even for the case of a singlet projection, for which the
simplified expression, Eq.~\eqref{eq:H_S}, can be utilized, producing
$\sim 3$ times more terms. In must be emphasized, however, that the
case of a singlet projection admits even more computationally
efficient solution within the constrained optimization approach
discussed by present authors in
Ref.~\citenum{Ryabinkin:2018/jctc/0-cnstr}. Namely, since $\hat S^2$
is a positively-defined operator, in other words,
$\braket{\Psi|\hat S^2|\Psi} \ge 0$ for any $\ket{\Psi}$, the solution
with $\braket{\Psi|\hat S^2|\Psi} = 0$ guarantees that $\ket{\Psi}$ is
the singlet eigenfunction of $\hat S^2$. Thus, if the constraining
minimization problem is feasible, the spin purity is guaranteed.
Finally, we note that the L\"owding projection technique was almost
always the worst solution; it is difficult to envision situations
where it could be recommended for a general use. However, it might be
used as a theoretical tool in finding qubit reduction schemes similar
to that proposed in Ref.~\citenum{Moll:2016/jpa/295301}.

\begin{acknowledgments}
  I.G.R. thanks Prof. Artur F. Izmaylov for numerous useful
  discussions.
\end{acknowledgments}

\appendix

\begingroup %
\renewcommand*{\arraystretch}{1.0}
\begin{table*}
  \centering
  \caption{Electronic energy levels (excluding nuclear-nuclear
    repulsion energy $V_{nn}$, in \si{\hartree}) along with their
    symmetry labels from diagonalization of the qubit Hamiltonian
    $\hat H$ for the \ce{LiH} molecule. The spectra of various
    number-projected operators are shown for comparison. The
    projection is done onto an $n = 2$ (a neutral molecule) subspace.
    The matched states are shown in bold.}
  \sisetup{%
    table-format = 3.6, %
    round-mode = places, %
    round-precision = 6, %
  } %
  \footnotesize
  \begin{tabularx}{0.60\linewidth}{@{}lcS|S|S|S@{}}
    \toprule
    Level & Symmetry label & \multicolumn{4}{Y}{Energy} \\\cmidrule{3-6}
    $k$   &  $(N_k, S_k)$  & \multicolumn{1}{c}{ $\hat H$}
          & \multicolumn{1}{c}{$\hat P \hat H \hat P$, Eq.\,\eqref{eq:projected_H}}
          & \multicolumn{1}{c}{$\hat{\mathcal{L}}$\footnotemark[1], Eq.\,\eqref{eq:symm_constr}}
          & \multicolumn{1}{c}{$\hat{\mathcal{H}}$, Eq.\,~\eqref{eq:H_A_decomp}} \\[0.5ex] 
    \midrule
    0  &  (2, 0.0)  &    \bfseries -8.2889385 & \bfseries -8.2889385  &\bfseries   -8.2889385  &\bfseries   -8.2889385 \\
    1  &  (2, 1.0)  &    \bfseries -8.2762330 & \bfseries -8.2762330  &\bfseries   -8.2762330  &\bfseries   -8.2762330 \\
    2  &  (2, 1.0)  &    \bfseries -8.2762330 & \bfseries -8.2762330  &\bfseries   -8.2762330  &\bfseries   -8.2762330 \\
    3  &  (2, 1.0)  &    \bfseries -8.2762330 & \bfseries -8.2762330  &\bfseries   -8.2762330  &\bfseries   -8.2762330 \\
    4  &  (2, 0.0)  &    \bfseries -8.2136207 & \bfseries -8.2136207  &\bfseries   -8.2136207  &\bfseries   -8.2136207 \\
    5  &  (3, 0.5)  &              -8.2037508 & \bfseries -8.1815168  &\bfseries   -8.1815168  &\bfseries   -8.1815168 \\
    6  &  (3, 0.5)  &              -8.2037508 & \bfseries -8.1815168  &\bfseries   -8.1815168  &\bfseries   -8.1815168 \\
    7  &  (2, 1.0)  &    \bfseries -8.1815168 & \bfseries -8.1815168  &\bfseries   -8.1815168  &\bfseries   -8.1815168 \\
    8  &  (2, 1.0)  &    \bfseries -8.1815168 & \bfseries -7.9036036  &\bfseries   -7.9036036  &\bfseries   -7.9036036 \\
    9  &  (2, 1.0)  &    \bfseries -8.1815168 & \bfseries -7.8793810  &\bfseries   -7.8793810  &\bfseries   -7.8793810 \\
    10 &  (3, 0.5)  &              -8.1636294 & \bfseries -7.8338657  &\bfseries   -7.8338657  &\bfseries   -7.8338657 \\
    11 &  (3, 0.5)  &              -8.1636294 & \bfseries -7.8338657  &\bfseries   -7.8338657  &\bfseries   -7.8338657 \\
    12 &  (3, 1.5)  &              -8.1298014 & \bfseries -7.8338657  &\bfseries   -7.8338657  &\bfseries   -7.8338657 \\
    13 &  (3, 1.5)  &              -8.1298014 & \bfseries -7.7438532  &\bfseries   -7.7438532  &\bfseries   -7.7438532 \\
    14 &  (3, 1.5)  &              -8.1298014 & \bfseries -7.6099413  &\bfseries   -7.6099413  &\bfseries   -7.6099413 \\
    15 &  (3, 1.5)  &              -8.1298014 &  -0.0000000  &  -0.2037508  &   7.4803138 \\
    16 &  (1, 0.5)  &              -8.1029489 &  -0.0000000  &  -0.2037508  &   7.4803138 \\
    17 &  (1, 0.5)  &              -8.1029489 &  -0.0000000  &  -0.1636294  &   7.6112886 \\
    18 &  (3, 0.5)  &              -8.0281577 &  -0.0000000  &  -0.1636294  &   7.6112886 \\
    19 &  (3, 0.5)  &              -8.0281577 &  -0.0000000  &  -0.1298014  &   7.7101065 \\
    20 &  (3, 0.5)  &              -7.9624971 &  -0.0000000  &  -0.1298014  &   7.7101065 \\
    21 &  (3, 0.5)  &              -7.9624971 &  -0.0000000  &  -0.1298014  &   7.7881749 \\
    22 &  (3, 0.5)  &              -7.9128128 &  -0.0000000  &  -0.1298014  &   7.7881749 \\
    23 &  (3, 0.5)  &              -7.9128128 &  -0.0000000  &  -0.1029489  &   7.8147873 \\
    24 &  (2, 0.0)  &    \bfseries -7.9036036 &  -0.0000000  &  -0.1029489  &   7.8147873 \\
    25 &  (2, 0.0)  &    \bfseries -7.8793810 &  -0.0000000  &  -0.0281577  &   7.9128128 \\
    26 &  (2, 1.0)  &    \bfseries -7.8338657 &  -0.0000000  &  -0.0281577  &   7.9128128 \\
    27 &  (2, 1.0)  &    \bfseries -7.8338657 &  -0.0000000  &   0.0375029  &   7.9624971 \\
    28 &  (2, 1.0)  &    \bfseries -7.8338657 &  -0.0000000  &   0.0375029  &   7.9624971 \\
    29 &  (4, 0.0)  &              -7.8304159 &  -0.0000000  &   0.0871872  &   8.0281577 \\
    30 &  (1, 0.5)  &              -7.8147873 &  -0.0000000  &   0.0871872  &   8.0281577 \\
    31 &  (1, 0.5)  &              -7.8147873 &  -0.0000000  &   0.1852127  &   8.1029489 \\
    32 &  (3, 0.5)  &              -7.7881749 &  -0.0000000  &   0.1852127  &   8.1029489 \\
    33 &  (3, 0.5)  &              -7.7881749 &  -0.0000000  &   0.2118251  &   8.1298014 \\
    34 &  (4, 1.0)  &              -7.7512655 &  -0.0000000  &   0.2118251  &   8.1298014 \\
    35 &  (4, 1.0)  &              -7.7512655 &  -0.0000000  &   0.2898935  &   8.1298014 \\
    36 &  (4, 1.0)  &              -7.7512655 &  -0.0000000  &   0.2898935  &   8.1298014 \\
    37 &  (2, 0.0)  &    \bfseries -7.7438532 &  -0.0000000  &   0.3887114  &   8.1636294 \\
    38 &  (1, 0.5)  &              -7.7101065 &  -0.0000000  &   0.3887114  &   8.1636294 \\
    39 &  (1, 0.5)  &              -7.7101065 &  -0.0000000  &   0.5196862  &   8.2037508 \\
    40 &  (4, 1.0)  &              -7.6768132 &  -0.0000000  &   0.5196862  &   8.2037508 \\
    41 &  (4, 1.0)  &              -7.6768132 &  -0.0000000  &  24.1695841  &  49.2173760 \\
    42 &  (4, 1.0)  &              -7.6768132 &  -0.0000000  &  24.2487345  &  51.5636357 \\
    43 &  (4, 0.0)  &              -7.6699445 &  -0.0000000  &  24.2487345  &  52.2621998 \\
    44 &  (4, 0.0)  &              -7.6252537 &  -0.0000000  &  24.2487345  &  52.4148559 \\
    45 &  (3, 0.5)  &              -7.6112886 &  -0.0000000  &  24.3231868  &  52.8446581 \\
    46 &  (3, 0.5)  &              -7.6112886 &  -0.0000000  &  24.3231868  &  52.8446581 \\
    47 &  (2, 0.0)  &    \bfseries -7.6099413 &  -0.0000000  &  24.3231868  &  52.8446581 \\
    48 &  (4, 1.0)  &              -7.5492369 &   0.0000000  &  24.3300555  &  53.3767759 \\
    49 &  (4, 1.0)  &              -7.5492369 &   0.0000000  &  24.3747463  &  53.6896112 \\
    50 &  (4, 1.0)  &              -7.5492369 &   0.0000000  &  24.4507631  &  53.7376921 \\
    51 &  (4, 0.0)  &              -7.4878366 &   0.0000000  &  24.4507631  &  53.7376921 \\
    52 &  (3, 0.5)  &              -7.4803138 &   0.0000000  &  24.4507631  &  53.7376922 \\
    53 &  (3, 0.5)  &              -7.4803138 &   0.0000000  &  24.5121634  &  54.2588582 \\
    54 &  (0, 0.0)  &              -7.4660285 &   0.0000000  &  24.5339715  &  54.2588582 \\
    55 &  (4, 0.0)  &              -7.3662337 &   0.0000000  &  24.6337663  &  54.2588582 \\
    56 &  (5, 0.5)  &              -7.1330137 &   0.0000000  &  24.9689463  &  54.8129115 \\
    57 &  (5, 0.5)  &              -7.1330137 &   0.0000000  &  64.8669863  & 117.4694243 \\
    58 &  (4, 0.0)  &              -7.0310537 &   0.0000000  &  64.8669863  & 117.4694243 \\
    59 &  (5, 0.5)  &              -6.9562628 &   0.0000000  &  65.0437372  & 118.2564669 \\
    60 &  (5, 0.5)  &              -6.9562628 &   0.0000000  &  65.0437372  & 118.2564669 \\
    61 &  (5, 0.5)  &              -6.9099661 &   0.0000000  &  65.0900339  & 121.2612329 \\
    62 &  (5, 0.5)  &              -6.9099661 &   0.0000000  &  65.0900339  & 121.2612329 \\
    63 &  (6, 0.0)  &              -6.1517285 &   0.0000000  & 121.8482715  & 190.7035829 \\
    \bottomrule
  \end{tabularx}
  \footnotetext[1]{Penalty parameter $\mu = \SI{15}{\hartree}$}
  \label{tab:spectra_comp}
\end{table*}
\endgroup

\end{document}